\documentclass[%
 aip,
 jmp,%
 amsmath,amssymb,
 reprint,%
]{revtex4-1}

\usepackage{graphicx}
\usepackage{subfigure}
\usepackage{dcolumn}
\usepackage{bm}
\usepackage{multirow}
\usepackage{algorithm}
\usepackage{algorithmic}

\usepackage{tabularx}
\usepackage{booktabs}
\usepackage{amssymb}
\usepackage{array}
\usepackage{epsfig}
\usepackage{epsf}

\begin{document}

\preprint{AIP/123-QED}

\title{A quantum extension to inspection game}

\author{Xinyang Deng}
\affiliation{School of Computer and Information Science, Southwest University, Chongqing 400715, China}
\affiliation{Center for Quantitative Sciences, Vanderbilt University School of Medicine, Nashville, TN 37232, USA}

\author{Yong Deng}%
\email[Corresponding author. E-mail: ]{ydeng@swu.edu.cn}
\affiliation{School of Computer and Information Science, Southwest University, Chongqing 400715, China}
\affiliation{School of Engineering, Vanderbilt University, Nashville, TN 37235, USA}%

\author{Qi Liu}%
\affiliation{Center for Quantitative Sciences, Vanderbilt University School of Medicine, Nashville, TN 37232, USA}
\affiliation{Department of Biomedical Informatics, Vanderbilt University School of Medicine, Nashville, TN 37232, USA}%

\author{Zhen Wang}%
\email[Corresponding author. E-mail: ]{zhenwang0@gmail.com}
\affiliation{School of Computer and Information Science, Southwest University, Chongqing 400715, China}
\affiliation{Interdisciplinary Graduate School of Engineering Sciences, Kyushu University, Kasuga-koen, Kasuga-shi, Fukuoka 816-8580, Japan}%

\date{\today}

\begin{abstract}
Quantum game theory is a new interdisciplinary field between game theory and physical research. In this paper, we extend the classical inspection game into a quantum game version by quantizing the strategy space and importing entanglement between players. Our result shows that the quantum inspection game has various Nash equilibrium depending on the initial quantum state of the game. It is also shown that quantization can respectively help each player to increase his own payoff, yet fails to bring Pareto improvement for the collective payoff in the quantum inspection game.
\end{abstract}

\keywords{Inspection game; Quantum game; Entanglement; Pareto improvement}

\maketitle

\begin{quotation}
Game theory is one of the most importance tools to research the human behavior and its potential on various myriad subjects. As a recently developing field of game theory, quantum games exhibit many new features which are greatly different from their classical counterparts. The Marinatto-Weber scheme provides a straightforward way to incorporate quantum viewpoints with classical game theory. In this paper, we quantize the inspection game based on the Marinatto-Weber scheme to explore whether the quantization could increase not only individual but also collective payoff, without decreasing the interests of others. The results show that this type of quantum treatment is just able to enhance each player's payoff, but can not realize a Pareto improvement.
\end{quotation}

\section{Introduction}
Game theory, founded by von Neumann and Morgenstern \cite{von2007theory} in 1940s, is a mathematical framework to explain and address the interactive decision situations, where the aims, goals and preferences of the participating agents are potentially in conflict. A solution to a game, which is self enforcing and where no player has an incentive to deviate from the present strategy unilaterally, forms the so-called Nash equilibrium (NE) \cite{nash1950equilibrium}. For example, in a two-person game, a combination of each player's strategy $(s_1^{*}, s_2^{*})$ is a NE if
\begin{equation}
\begin{array}{l}
u_1 (s_1^* ,s_2^* ) \ge u_1 (s_1 ,s_2^* ),\quad \forall s_1  \in S_1, \\
u_2 (s_1^* ,s_2^* ) \ge u_2 (s_1^* ,s_2 ),\quad \forall s_2  \in S_2, \\
\end{array}
\end{equation}
where the payoffs of two players are determined by functions $u_1(s_1, s_2)$ and $u_2(s_1, s_2)$, respectively. Through the development of several decades, game theory has gradually become one of the most importance tools to research the human behavior and its potential on various myriad subjects \cite{szolnoki2009topology,mobilia2012stochastic,poncela2009evolutionary,hui2007cooperation,chen2008interaction,wang2015ScalingPLR,gracia2014intergroup,wang2015evolutionaryEPJB,bauch2004vaccination,deng2014belief,deng2014impact,santos2005scale,rulquin2014globally,jusup2014barriers,wang2013insight,wang2013impact,boccaletti2014structure}.

Recently, an important breakthrough of game theory is closely related with quantum information theory \cite{klarreich2001playing}. In particular, physicists have explored the quantization of games and presented quantum games and quantum strategies. For example, when  classical framework of games was incorporated with the quantum viewpoints, Eisert {\em et al.} used quantum approaches to deal with the prisoner's dilemma and realized Pareto efficiency where
it is impossible to make one party better off without making someone worse off~\cite{eisert1999quantum}.  Meyer found that a player who implemented a quantum strategy could increase his expected payoff by using a PQ penny flipover game~\cite{meyer1999quantum}. Furthermore, a lot of research works have proven that quantum games are greatly different from their classical counterparts \cite{marinatto2000quantum,iqbal2001evolutionarily,bo2003quantum,du2002entanglement,benjamin2001multiplayer,
li2012evolutionEPJB,situ2014quantum,pawela2013quantum,li2014entanglement}, which thus indicates that quantum games open up a new way to study the dilemmas in the classical game theory.

In this paper, we investigate the inspection game, and extend it into a quantum version. The inspection game \cite{dresher1962sampling} is a useful metaphor to describe the relationship between two individuals who have conflicting interest \cite{kamal2015evolutionary}. In its basic version, one player chooses to inspect or not, the other chooses to comply or not. Since both players have conflict of interest, the game has a mixed-strategy NE that is not a Pareto efficiency. Here we quantize the inspection game to explore whether the quantization could increase not only individual but also collective payoff, without decreasing the interests of others. The results show that this type of quantum treatment is just able to enhance each player's payoff, but can not realize a Pareto improvement.

\section{Inspection game}
Inspection game was first proposed by Dresher \cite{dresher1962sampling} to describe the interactions between two agents with conflicts of interest, such as inspector and smuggler, employer and worker. Following generalization description of inspection game \cite{fudenberg1991game,nosenzo2014encouraging,kamal2015evolutionary}, there are an employer, who can either inspect (\emph{I}) or not (\emph{N}), and a worker, who can either work (\emph{W}) or shirk (\emph{S}). As illustrated in the payoff matrix (see Eq.~(\ref{IGPayoffMatrix})), the row (column) player denotes employer (worker).
The employer bears a cost of $h$ from inspecting, and pays the worker a wage of $w$ unless he finds that the worker is shirking. The worker bears a work-related cost $g$, and creates an wealth $v$ for the employer if he works. It is assumed that all variables are positive and $v>g$, $w>h$, $w>g$. Obviously, the joint payoff is maximized when the worker works and the employer does not inspect.

\begin{equation}\label{IGPayoffMatrix}
\begin{array}{*{20}c}
   {} & {\begin{array}{*{20}c}
   \quad W & {\quad \qquad \qquad S}  \\
\end{array}}  \\
   {\begin{array}{*{20}c}
   I  \\
   N  \\
\end{array}} & {\left( {\begin{array}{*{20}c}
   {(v - w - h,w - g)} & {( - h,0)}  \\
   {(v - w,w - g)} & {( - w,w)}  \\
\end{array}} \right).}  \\
\end{array}
\end{equation}

For the above payoff matrix (Eq.~\ref{IGPayoffMatrix}), there does not exist a pure-strategy NE due to $v - w > v - w - h$, $-h > -w$, $w-g > 0$ and $w > w-g$. But it has a mixed-strategy NE where the employer inspects with probability $p=g/w$ and the worker works with probability $q=1-h/w$. In this equilibrium the employer gets a payoff  $\bar \$_{A}= v-w-hv/w$, the worker receives a payoff  $\bar \$_{B}= w-g$, and the joint payoff is $\bar \$_{A+B}= v-c-hv/w$. To illustrate this point, let us consider the following example \cite{nosenzo2014encouraging}:
\begin{equation}
v = 60,g = 15,h = 8,w = 20.
\end{equation}
Then the payoff matrix (Eq.~\ref{IGPayoffMatrix}) takes the form
\begin{equation}\label{SpecificIGPayoffMatrix}
\begin{array}{*{20}c}
   {} & {\begin{array}{*{20}c}
   W & \qquad S  \\
\end{array}}  \\
   {\begin{array}{*{20}c}
   I  \\
   N  \\
\end{array}} & {\left( {\begin{array}{*{20}c}
   {(32,5)} & {( - 8,0)}  \\
   {(40,5)} & {( - 20,20)}  \\
\end{array}} \right).}  \\
\end{array}
\end{equation}
It is clear that there is no pure-strategy NE in this game. Now let us check its mixed-strategy NE. We assume that the employer inspects with probability $p$ and the worker works with probability $q$. The expected payoff of each player is defined as
\begin{equation}
\begin{array}{l}
\bar \$_i (p,q)  = pqE_i (I,W) + p(1 - q)E_i (I,S) \\
 \qquad + (1 - p)qE_i (N,W) + (1 - p)(1 - q)E_i (N,S), \\
\end{array}
\end{equation}
where $E_i(s_1, s_2)$ is player $i$'s payoff for strategy $s_1$ against strategy $s_2$ of his opponent. So, for employer $A$ and worker $B$, their expected payoffs are
\begin{equation}
\left\{ \begin{array}{l}
 \bar \$ _A (p,q) = 12p + 60q - 20pq - 20, \\
 \bar \$ _B (p,q) = 20pq - 15q - 20p + 20. \\
 \end{array} \right.
\end{equation}
If there does exist a mixed-strategy NE, this equilibrium can be found by calculating $\frac{{\partial \bar \$ _A (p,q)}}{{\partial p}} = 0$ and $\frac{{\partial \bar \$ _B (p,q)}}{{\partial q}} = 0$. The results are $p^* = 0.75$, $q^*=0.6$. It is easy to demonstrate that $(p^*,q^*)$ is indeed a mixed-strategy NE because
\begin{equation}
\left\{ \begin{array}{l}
 \bar \$ _A (p^* ,q^* ) \ge \bar \$ _A (p,q^* ),\quad \forall p \in [0,1], \\
 \bar \$ _B (p^* ,q^* ) \ge \bar \$ _B (p^* ,q),\quad \forall q \in [0,1]. \\
 \end{array} \right.
\end{equation}
In this NE, the expected payoffs of employer $A$ and worker $B$ are $\bar \$ _A^* = 16$, $\bar \$ _B^* = 5$, respectively. And the joint payoff is $\bar \$ _{A+B}^*=21$.

\section{Quantum inspection game}
Up to the present, there have been several scenarios, for example Eisert-Wilkens-Lewenstein scheme \cite{eisert1999quantum} and Marinatto-Weber scheme \cite{marinatto2000quantum}, to quantize the classical strategy space. In this section, we will follow Marinatto-Weber version to quantize the strategy space for inspection game.

Let us define a four-dimensional Hilbert space $H$ for the inspection game by giving its orthonormal basis vectors $H = H_A \otimes H_B = \{ \left| {IW} \right\rangle, \left| {IS} \right\rangle, \left| {NW} \right\rangle, \left| {NS} \right\rangle \}$, where the first qubit is reserved for the state of employer $A$ and the second one for that of worker $B$. We assume that these two players, employer $A$ and worker $B$, share the following quantum state:
\begin{equation}\label{PsiInABCD}
\left| {\psi _{in} } \right\rangle  = a\left| {IW} \right\rangle  + b\left| {IS} \right\rangle  + c\left| {NW} \right\rangle  + d\left| {NS} \right\rangle,
\end{equation}
where $|a|^2 + |b|^2 + |c|^2 + |d|^2 = 1$. According to state vector $\left| {\psi _{in} } \right\rangle$, the associated density matrix can be derived as $\rho _{in}  = \left| {\psi _{in} } \right\rangle \left\langle {\psi _{in} } \right|$.

Let $C$ be a unitary and Hermitian operator (i.e.,  such that $C^\dag   = C = C^{ - 1}$), such that
\begin{equation}
C\left| I \right\rangle = \left| N \right\rangle, C\left| N \right\rangle = \left| I \right\rangle, C\left| W \right\rangle = \left| S \right\rangle, C\left| S \right\rangle = \left| W \right\rangle.
\end{equation}
In the interaction process, employer $A$ does nothing with probability $p$ by using identity operator $I$ , and acts as operator $C$ on the first qubit with probability $1-p$. Analogously, worker $B$ does nothing with probability $q$ and performs $C$ on the second qubit with probability $1-q$. Then, the final density matrix for this two-qubit quantum system takes the form
\begin{equation}
\begin{array}{l}
\rho_{fin}   = pq\left[ {(I_A  \otimes I_B )\rho _{in} (I_A^\dag   \otimes I_B^\dag  )} \right] \\
 \quad \quad \quad  + p(1 - q)\left[ {(I_A  \otimes C_B )\rho _{in} (I_A^\dag   \otimes C_B^\dag  )} \right] \\
 \quad \quad \quad  + (1 - p)q\left[ {(C_A  \otimes I_B )\rho _{in} (C_A^\dag   \otimes I_B^\dag  )} \right] \\
 \quad \quad \quad  + (1 - p)(1 - q)\left[ {(C_A  \otimes C_B )\rho _{in} (C_A^\dag   \otimes C_B^\dag  )} \right] .\\
 \end{array}
\end{equation}

In order to calculate the payoff, two payoff operators are introduced as
\begin{equation}
\begin{array}{l}
 P_A  = E_A (I,W)\left| {IW} \right\rangle \left\langle {IW} \right| + E_A (I,S)\left| {IS} \right\rangle \left\langle {IS} \right| \\
 \quad \quad  + E_A (N,W)\left| {NW} \right\rangle \left\langle {NW} \right| + E_A (N,S)\left| {NS} \right\rangle \left\langle {NS} \right|, \\
 \end{array}
\end{equation}
\begin{equation}
\begin{array}{l}
 P_B  = E_B (I,W)\left| {IW} \right\rangle \left\langle {IW} \right| + E_B (I,S)\left| {IS} \right\rangle \left\langle {IS} \right| \\
 \quad \quad  + E_B (N,W)\left| {NW} \right\rangle \left\langle {NW} \right| + E_B (N,S)\left| {NS} \right\rangle \left\langle {NS} \right|. \\
 \end{array}
\end{equation}
Finally, the payoff functions of $A$ and $B$ can be obtained as mean values of these operators
\begin{equation}\label{PayoffFunctionsTrAB}
\bar \$ _A (p,q) = Tr(P_A \rho _{fin} ),\quad \bar \$ _B (p,q) = Tr(P_B \rho _{fin} ).
\end{equation}

As shown above, we successfully build quantum inspection game based on Marinatto-Weber scheme. Next, it is of interest to inspect the impact of quantization on the inspection game. For simplicity, we specify the payoff matrix of inspection game as Eq.~(\ref{SpecificIGPayoffMatrix}), which has been considered in the above section.

According to Eqs.~(\ref{SpecificIGPayoffMatrix}) and  (\ref{PsiInABCD})-(\ref{PayoffFunctionsTrAB}), the expected payoff functions for both players can be rewritten as follows
\begin{equation}
\begin{array}{l}
 \bar \$ _A (p ,q ) = p[q(20|b|^2  + 20|c|^2  - 20|a|^2  - 20|d|^2 ) \\
 \quad \quad \quad \quad  + (12|a|^2  - 12|c|^2  + 8|d|^2  - 8|b|^2 )] \\
 \quad \quad \quad \quad  + q(60|a|^2  - 60|b|^2  + 40|c|^2  - 40|d|^2 ) \\
 \quad \quad \quad \quad  + (40|b|^2  - 20|a|^2  - 8|c|^2  + 32|d|^2 ), \\
 \bar \$ _B (p,q) = q[p(20|a|^2  - 20|b|^2  - 20|c|^2  + 20|d|^2 ) \\
 \quad \quad \quad \quad  + (15|b|^2  - 15|a|^2  + 5|c|^2  - 5|d|^2 )] \\
 \quad \quad \quad \quad  + p(20|c|^2  - 20|a|^2 ) \\
 \quad \quad \quad \quad  + (20|a|^2  + 5|b|^2  + 5|d|^2 ). \\
 \end{array}
\end{equation}

A NE $(p^*, q^*)$ can be obtained by imposing the following two conditions:
\begin{equation}\label{EqNEconditions}
\begin{array}{l}
 \bar \$ _A (p^* ,q^* ) - \bar \$ _A (p,q^* ) = \\
 \qquad  (p^*  - p)[q^* (20|b|^2  + 20|c|^2  - 20|a|^2  - 20|d|^2 ) \\
 \qquad   + (12|a|^2  - 12|c|^2  + 8|d|^2  - 8|b|^2 )] \ge 0, \;\; \forall p \in [0,1], \\
 \bar \$ _B (p^* ,q^* ) - \bar \$ _B (p^* ,q) = \\
 \qquad (q^*  - q)[p^* (20|a|^2  - 20|b|^2  - 20|c|^2  + 20|d|^2 ) \\
 \quad   + (15|b|^2  - 15|a|^2  + 5|c|^2  - 5|d|^2 )] \ge 0, \;\; \forall q \in [0,1]. \\
 \end{array}
\end{equation}

According to inequalities (\ref{EqNEconditions}), we can find the required condition for any possible NE. There are five cases.

{\textbf{Case 1: let $(p^*=1, q^*=1)$ be a NE. }}
Let us consider a case of $(p^*=1, q^*=1)$ and find the conditions of it being a NE. In such a case, the conditions shown in inequalities (\ref{EqNEconditions}) translate to
\begin{equation}\label{EqNEconditionsCase11}
\begin{array}{l}
 \bar \$ _A (1,1) - \bar \$ _A (p,1) = \\
 (1 - p)(12|b|^2  + 8|c|^2  - 8|a|^2  - 12|d|^2 ) \ge 0,\;\; \forall p \in [0,1], \\
 \bar \$ _B (1,1) - \bar \$ _B (1,q) = \\
 (1 - q)(5|a|^2  - 5|b|^2  - 15|c|^2  + 15|d|^2 ) \ge 0,\;\; \forall q \in [0,1]. \\
 \end{array}
\end{equation}

The inequalities (\ref{EqNEconditionsCase11}) require
\begin{equation}
C1 \quad \left\{ \begin{array}{l}
 - 2|a|^2 + 3|b|^2  + 2|c|^2  - 3|d|^2  \ge 0, \\
 |a|^2  - |b|^2  - 3|c|^2  + 3|d|^2  \ge 0, \\
 |a|^2  + |b|^2  + |c|^2  + |d|^2  = 1. \\
 \end{array} \right.
\end{equation}

The above conditions $C1$ holds for example, when $|a|^2  = 0.6,|b|^2  = 0.4,|c|^2  = |d|^2  = 0$. In this case of $(p^*=1, q^*=1)$ being a NE, the corresponding payoff functions are found as follows:
\begin{equation}\label{EqPayoffNEp1q1}
\left\{ \begin{array}{l}
 \bar \$ _A (p^*  = 1,q^* =1) = 32|a|^2  - 8|b|^2  + 40|c|^2  - 20|d|^2,  \\
 \bar \$ _B (p^*  = 1,q^* =1) = 5|a|^2  + 5|c|^2  + 20|d|^2.  \\
 \end{array} \right.
\end{equation}

According to condition $C1$ and Eqs.~(\ref{EqPayoffNEp1q1}), the range of each payoff can be found easily.
\begin{equation}\label{EqNEp1q1PayoffsRanges}
\left\{ \begin{array}{l}
 \bar \$ _A (p^*  = 1,q^*  = 1) \in [ - 14,19.333], \\
 \bar \$ _B (p^*  = 1,q^*  = 1) \in [2.5,10.625], \\
 \bar \$ _{A + B} (p^*  = 1,q^*  = 1) \in [ - 6,22.667]. \\
 \end{array} \right.
\end{equation}
These results show that either the payoff of each player or two players' joint payoff could increase in the quantum inspection game. Moreover, from Eqs.~(\ref{EqNEp1q1PayoffsRanges}), the maximum joint payoff is 22.667. Recalling the classical inspection game, we find that it has a unique NE where $\bar \$ _A^* = 16$, $\bar \$ _B^* = 5$ and $\bar \$ _{A+B}^*=21$. A key question becomes whether the increase of joint payoff in quantum inspection game results from the simultaneous increase of each player's payoff. If yes, the quantum version of inspection game definitely outperforms the classical one because it carries out a Pareto improvement. We give the following optimization problem to answer this question:
\begin{equation}\label{Case11ParetoImprovement}
\begin{array}{l}
 \max \quad \bar \$ _{A + B} (p^*  = 1,q^*  = 1) = \\
  \qquad \qquad \bar \$ _A (p^*  = 1,q^*  = 1) + \bar \$ _B (p^*  = 1,q^*  = 1) \\
 s.t.\quad \left\{ \begin{array}{l}
 \bar \$ _A (p^*  = 1,q^*  = 1) \ge 16, \\
 \bar \$ _B (p^*  = 1,q^*  = 1) \ge 5, \\
 - 2|a|^2  + 3|b|^2  + 2|c|^2  - 3|d|^2  \ge 0, \\
 |a|^2  - |b|^2  - 3|c|^2  + 3|d|^2  \ge 0, \\
 |a|^2  + |b|^2  + |c|^2  + |d|^2  = 1. \\
 \end{array} \right. \\
 \end{array}
\end{equation}
For the optimization problem of Eq.~(\ref{Case11ParetoImprovement}), the maximum of object function is 21 when $|a|^2  = 0.45$, $|b|^2  = 0.3$, $|c|^2  = 0.15$, $|d|^2  = 0.1$, at that situation $\$ _A (p^*  = 1,q^*  = 1) = 16$ and $\bar \$ _B (p^*  = 1,q^*  = 1) = 5$. Therefore, in this case, the quantum inspection game increases either the payoff of employer $A$ or the payoff of worker $B$, but can not simultaneously increase the payoffs of $A$ and $B$, compared with the equilibrium of classical inspection game. That is to say, the quantum inspection game does not carry out a Pareto improvement.

{\textbf{Case 2: let $(p^*=0, q^*=0)$ be a NE. }}
Let us examine another case of $(p^*=0, q^*=0)$ and find the conditions of it being a NE. In such a case, the conditions shown in inequalities (\ref{EqNEconditions}) translate to
\begin{equation}\label{EqNEconditionsCasep0q0}
\begin{array}{l}
 \bar \$ _A (0,0) - \bar \$ _A (p,0) = \\
 \quad p(12|c|^2  - 12|a|^2  + 8|b|^2  - 8|d|^2 ) \ge 0,\quad \forall p \in [0,1], \\
 \bar \$ _B (0,0) - \bar \$ _B (0,q) = \\
 \quad q(15|a|^2  - 15|b|^2  - 5|c|^2  + 5|d|^2 ) \ge 0,\quad \forall q \in [0,1]. \\
 \end{array}
\end{equation}

The inequalities (\ref{EqNEconditionsCasep0q0}) require
\begin{equation}
C2 \quad \left\{ \begin{array}{l}
  - 3|a|^2  + 2|b|^2  + 3|c|^2  - 2|d|^2  \ge 0, \\
 3|a|^2  - 3|b|^2  - |c|^2  + |d|^2  \ge 0, \\
 |a|^2  + |b|^2  + |c|^2  + |d|^2  = 1. \\
 \end{array} \right.
 \end{equation}

The above conditions $C2$ hold, for example, when $|a|^2  = 0.4375$, $|b|^2  = 0.375$, $|c|^2  = 0.1875$, $|d|^2  = 0$. In this case of $(p^*=0, q^*=0)$ being a NE, the corresponding payoff functions are found as follows:
\begin{equation}\label{EqPayoffNEp0q0}
\left\{ \begin{array}{l}
 \bar \$ _A (p^* = 0,q^* = 0) = 40|b|^2 - 20|a|^2  - 8|c|^2  + 32|d|^2, \\
 \bar \$ _B (p^* = 0,q^* = 0) = 20|a|^2  + 5|b|^2  + 5|d|^2.  \\
 \end{array} \right.
\end{equation}

According to conditions $C2$ and Eqs.~(\ref{EqPayoffNEp0q0}), the range of each payoff in this case is
\begin{equation}\label{EqNEp0q0PayoffsRanges}
\left\{ \begin{array}{l}
 \bar \$ _A (p^*  = 0,q^*  = 0) \in [ -14, 19.333], \\
 \bar \$ _B (p^*  = 0,q^*  = 0) \in [2.5, 10.625], \\
 \bar \$ _{A + B} (p^*  = 0,q^*  = 0) \in [ -6, 22.667]. \\
 \end{array} \right.
\end{equation}

These results are identical with the case of $(p^*  = 1,q^*  = 1)$. Similarly, we examine if there exists possible Pareto improvement based on the following optimization equation
\begin{equation}\label{Casep0q0ParetoImprovement}
\begin{array}{l}
 \max \quad \bar \$ _{A + B} (p^*  = 0,q^*  = 0) = \\
 \qquad \qquad \bar \$ _A (p^*  = 0,q^*  = 0) + \bar \$ _B (p^*  = 0,q^*  = 0) \\
 s.t.\quad \left\{ \begin{array}{l}
 \bar \$ _A (p^*  = 0,q^*  = 0) \ge 16, \\
 \bar \$ _B (p^*  = 0,q^*  = 0) \ge 5, \\
  - 3|a|^2  + 2|b|^2  + 3|c|^2  - 2|d|^2  \ge 0, \\
 3|a|^2  - 3|b|^2  - |c|^2  + |d|^2  \ge 0, \\
 |a|^2  + |b|^2  + |c|^2  + |d|^2  = 1. \\
 \end{array} \right. \\
 \end{array}
\end{equation}
For Eq.~(\ref{Casep0q0ParetoImprovement}), the maximum of object function is 21 when $|a|^2  = 0.1$, $|b|^2  = 0.15$, $|c|^2  = 0.3$, $|d|^2  = 0.45$, at that situation $\$ _A (p^*  = 0,q^*  = 0) = 16$ and $\bar \$ _B (p^*  = 0,q^*  = 0) = 5$. Therefore, in this case of $(p^*  = 0, q^*  = 0)$ being a NE, the quantum inspection game still does not carry out a Pareto improvement compared with the NE of classical inspection game.

{\textbf{Case 3: let $(p^*=1, q^*=0)$ be a NE. }}
Now let us consider the case of $(p^*=1, q^*=0)$ to find the conditions of it being a NE. In this case, the conditions shown in inequalities (\ref{EqNEconditions}) translate to
\begin{equation}\label{EqNEconditionsCasep1q0}
\begin{array}{l}
 \bar \$ _A (1,0) - \bar \$ _A (p,0) = \\
 \;\; (1 - p)(12|a|^2  - 8|b|^2  - 12|c|^2  + 8|d|^2 ) \ge 0, \forall p \in [0,1], \\
 \bar \$ _B (1,0) - \bar \$ _B (1,q) = \\
 \;\; q(-5|a|^2  + 5|b|^2  + 15|c|^2  - 15|d|^2 ) \ge 0,\quad \forall q \in [0,1]. \\
 \end{array}
\end{equation}

The inequalities (\ref{EqNEconditionsCasep1q0}) require
\begin{equation}
C3 \quad \left\{ \begin{array}{l}
 3|a|^2  - 2|b|^2  - 3|c|^2  + 2|d|^2  \ge 0, \\
  -|a|^2  + |b|^2  + 3|c|^2  - 3|d|^2  \ge 0. \\
 |a|^2  + |b|^2  + |c|^2  + |d|^2  = 1. \\
 \end{array} \right.
\end{equation}

Conditions $C3$ hold, for example, when $|a|^2  = 0.75$, $|b|^2  = 0$, $|c|^2  = 0.25$, $|d|^2  = 0$. In this case of $(p^*=1, q^*=0)$ being a NE, the corresponding payoff functions are shown as follows:
\begin{equation}\label{EqPayoffNEp1q0}
\left\{ \begin{array}{l}
 \bar\$_A (p^*  = 1,q^*  = 0) = 32|b|^2 - 8|a|^2  - 20|c|^2  + 40|d|^2,\\
 \bar\$_B (p^*  = 1,q^*  = 0) = 5|b|^2  + 20|c|^2  + 5|d|^2.\\
 \end{array} \right.
\end{equation}

Based on condition $C3$ and Eqs.~(\ref{EqPayoffNEp1q0}), we can verify that the range of each payoff is completely same with that of above cases, namely
\begin{equation}\label{EqNEp1q0PayoffsRanges}
\left\{ \begin{array}{l}
 \bar \$ _A (p^*  = 1,q^*  = 0) \in [ -14, 19.333], \\
 \bar \$ _B (p^*  = 1,q^*  = 0) \in [2.5, 10.625], \\
 \bar \$ _{A + B} (p^*  = 1,q^*  = 0) \in [ -6, 22.667]. \\
 \end{array} \right.
\end{equation}

Similarly, we find that in this case of $(p^*  = 1,q^*  = 0)$ being a NE, the quantum version of inspection game is still not able to implement a Pareto improvement on the mixed-strategy NE of classical inspection game. The maximum joint payoff of $A$ and $B$ is 21 when $|a|^2  = 0.3$, $|b|^2  = 0.45$, $|c|^2  = 0.1$, $|d|^2  = 0.15$, at the same time $\$ _A (p^*  = 1,q^*  = 0) = 16$ and $\bar \$ _B (p^*  = 1,q^*  = 0) = 5$.

{\textbf{Case 4: let $(p^*=0, q^*=1)$ be a NE. }}
Let us explore the case of $(p^*=0, q^*=1)$ being a NE. In such a case, the conditions shown in inequalities (\ref{EqNEconditions}) translate to
\begin{equation}\label{EqNEconditionsCasep0q1}
\begin{array}{l}
 \bar \$ _A (0,1) - \bar \$ _A (p,1) = \\
 \;  p(8|a|^2  - 12|b|^2  - 8|c|^2  + 12|d|^2 ) \ge 0,\quad \forall p \in [0,1], \\
 \bar \$ _B (0,1) - \bar \$ _B (0,q) = \\
  \; (1 - q)( 15|b|^2 - 15|a|^2  + 5|c|^2  - 5|d|^2 ) \ge 0, \forall q \in [0,1]. \\
 \end{array}
\end{equation}

The inequalities (\ref{EqNEconditionsCasep0q1}) require
\begin{equation}
C4 \quad \left\{ \begin{array}{l}
 2|a|^2  - 3|b|^2  - 2|c|^2  + 3|d|^2  \ge 0, \\
  - 3|a|^2  + 3|b|^2  + |c|^2  - |d|^2  \ge 0, \\
 |a|^2  + |b|^2  + |c|^2  + |d|^2  = 1. \\
 \end{array} \right.
\end{equation}

The condition $C4$ holds, for example, when $|a|^2  = |c|^2  = 0$, $|b|^2  = |d|^2  = 0.5$. In this case of $(p^*=0, q^*=1)$ being a NE, the corresponding payoff functions are shown as follows:
\begin{equation}\label{EqPayoffNEp0q1}
\left\{ \begin{array}{l}
 \bar \$ _A (p^*  = 0,q^*  = 1) = 40|a|^2  - 20|b|^2  + 32|c|^2  - 8|d|^2,  \\
 \bar \$ _B (p^*  = 0,q^*  = 1) = 5|a|^2  + 20|b|^2  + 5|c|^2.  \\
 \end{array} \right.
\end{equation}

Depending on condition $C4$ and Eqs.~(\ref{EqPayoffNEp0q1}), we can obtain the range of each payoff
\begin{equation}\label{EqNEp0q1PayoffsRanges}
\left\{ \begin{array}{l}
 \bar \$ _A (p^*  = 0,q^*  = 1) \in [ -14, 19.333], \\
 \bar \$ _B (p^*  = 0,q^*  = 1) \in [2.5, 10.625], \\
 \bar \$ _{A + B} (p^*  = 0,q^*  = 1) \in [ -6, 22.667]. \\
 \end{array} \right.
\end{equation}

In this case, we  find that there is still not Pareto improvement even adopting the quantum inspection game. The maximum joint payoff of $A$ and $B$, which equals to 21, is calculated when $|a|^2  = 0.15$, $|b|^2  = 0.1$, $|c|^2  = 0.45$, $|d|^2  = 0.3$, and $\$ _A (p^*  = 1,q^*  = 0) = 16$, $\bar \$ _B (p^*  = 1,q^*  = 0) = 5$.

{\textbf{Case 5: let $(p^*, q^*)$ be a mixed-strategy NE. }}
In this case, we first consider the situation of $p^*, q^* \in (0,1)$. For $p^*$ and $q^*$ differing from 0 or 1, the inequalities (\ref{EqNEconditions}) are required to be satisfied. Since the factors $(p^* - p)$ and $(q^* - q)$ may be positive or negative for different values of $p$ and $q$, the only way of fulfilling conditions shown in inequalities (\ref{EqNEconditions}) is to make the coefficients of $(p^* - p)$ and $(q^* - q)$ become zero. So, the NE can be obtained as follows
\begin{equation}
\begin{array}{l}
p^*  = \frac{{3|a|^2  - 3|b|^2  - |c|^2  + |d|^2 }}{{4|a|^2  - 4|b|^2  - 4|c|^2  + 4|d|^2 }}, \\
q^*  = \frac{{3|a|^2  - 2|b|^2  - 3|c|^2  + 2|d|^2 }}{{5|a|^2  - 5|b|^2  - 5|c|^2  + 5|d|^2 }}. \\
\end{array}
\end{equation}

The above mixed-strategy NE is constrained by several self-evident conditions denoted as $C5$: (i) $p^*, q^* \in (0,1)$; (ii) $|a|^2 + |b|^2  + |c|^2  + |d|^2 = 1$. According to the pair of $p^*$ and $q^*$, the payoff functions become
\begin{equation}
\left\{ \begin{array}{l}
 \bar \$ _A (p^* ,q^* ) = \\
 \qquad \qquad \frac{{16|a|^4  - 16|b|^4  - 16|c|^4  + 16|d|^4  + 12|a|^2 |d|^2  - 12|b|^2 |c|^2 }}{{|a|^2  - |b|^2  - |c|^2  + |d|^2 }}, \\
 \bar \$ _B (p^* ,q^* ) = \\
 \qquad \qquad \frac{{5|a|^4  - 5|b|^4  - 5|c|^4  + 5|d|^4  + 20|a|^2 |d|^2  - 20|b|^2 |c|^2 }}{{|a|^2  - |b|^2  - |c|^2  + |d|^2 }}. \\
 \end{array} \right.
\end{equation}

Likewise, the bound of each payoff is found as
\begin{equation}
\left\{ \begin{array}{l}
 \bar \$ _A (p^* ,q^* ) \in [11,16], \\
 \bar \$ _B (p^* ,q^* ) \in [5,7.5], \\
 \bar \$ _{A + B} (p^* ,q^* ) \in [18.5,21]. \\
 \end{array} \right.
\end{equation}

It is clear that there is not Pareto improvement in the quantum inspection game because the maximin joint payoff of $A$ and $B$ does not exceed 21. On the contrary, in this case the import of quantization decreases the collective payoff of all players. Moreover, the payoff of employer $A$ never exceeds 16, and payoff of worker $B$ is always above 5. Different from above mentioned cases, the present case (namely, $(p^*, q^*)$ being a NE where $p^*, q^* \in (0,1)$) is definitely beneficial for worker $B$ and harmful for employer $A$, compared with the classical inspection game.

In addition, a mixed-strategy NE may also be $(p^*, 0)$, $(p^*, 1)$, $(0, q^*)$ or $(1, q^*)$, where $p^*, q^* \in (0,1)$. For these cases, we have examined these possibilities one by one, and still find that Pareto improvement did not occur.

\section{Nash equilibriums of some typical quantum states}
To illustrate the features of quantum inspection game, some typical quantum states need to be set as initial states. In this section, we will provided some examples, which are helpful to further understand the quantum version of inspective game.

{\textbf{Example 1: $\left| {\psi _{in} } \right\rangle  = \left| {IW} \right\rangle$. }}

Let $|a|^2 = 1$, $|b|^2 = |c|^2 = |d|^2 = 0$, a simple initial state $\left| {\psi _{in} } \right\rangle  = \left| {IW} \right\rangle$ is obtained. According to quantum mechanics definition, this state is a factorizable quantum state. Based on Eqs (\ref{PsiInABCD}) - (\ref{EqNEconditions}), the NE of this game is readily found: it has a unique NE $(p^* = 0.75, q^* = 0.6)$; and the payoff of each player is $\bar \$ _A (0.75, 0.6) = 16$, $\bar \$ _B (0.75, 0.6) = 5$, which are identical with the outcomes of classical inspection game. That is to say,  the quantum game has reproduced the results of classical game theory in the case of factorizable quantum state.

{\textbf{Example 2: $\left| {\psi _{in} } \right\rangle  = \sqrt {\frac{1}{2}} \left| {IW} \right\rangle  + \sqrt {\frac{1}{2}} \left| {NS} \right\rangle$ or $\left| {\psi _{in} } \right\rangle  = \sqrt {\frac{1}{2}} \left| {IS} \right\rangle  + \sqrt {\frac{1}{2}} \left| {NW} \right\rangle$. }}

These two quantum states $\left| {\psi _{in} } \right\rangle  = \sqrt {\frac{1}{2}} \left| {IW} \right\rangle  + \sqrt {\frac{1}{2}} \left| {NS} \right\rangle$ and $\left| {\psi _{in} } \right\rangle  = \sqrt {\frac{1}{2}} \left| {IS} \right\rangle  + \sqrt {\frac{1}{2}} \left| {NW} \right\rangle$ are known as Bell states in quantum mechanics, showing that these two players are entangled in the initial states. In this case, the quantum inspection game has a same and unique NE  $(p^* = 0.5, q^* = 0.5)$, where leads to  $\bar \$ _A (0.5, 0.5) = 11$, $\bar \$ _B (0.5, 0.5) = 7.5$. These results show that worker $B$ benefits from the entanglement, but employer $A$ does not. Moreover, the entanglement decreases the joint payoff of $A$ and $B$.

{\textbf{Example 3: $\left| {\psi _{in} } \right\rangle  = \sqrt {\frac{1}{2}} \left| {IW} \right\rangle  + \sqrt {\frac{1}{2}} \left| {IS} \right\rangle$. }}

In this example, the quantum inspection game has countless NEs denoted as $\{ (1,q^* )|q^*  \in [0,1]\}$. Payoffs of players are $\bar \$ _A (1,q^* ) = 12$, $\bar \$ _B (1,q^* ) = 2.5$, $\forall q^* \in [0,1]$. Compared with the NE of classical inspection game, both  players' payoffs have decreased.

{\textbf{Example 4: $\left| {\psi _{in} } \right\rangle  = \sqrt {\frac{1}{2}} \left| {NW} \right\rangle  + \sqrt {\frac{1}{2}} \left| {NS} \right\rangle$. }}

The results of this example are given as follows, which is similar to Example 3.

NEs: $\{ (0,q^* )|q^*  \in [0,1]\}$.

$\bar \$ _A (0,q^* ) = 12$, $\bar \$ _B (0,q^* ) = 2.5$, $\forall q^* \in [0,1]$.

{\textbf{Example 5: $\left| {\psi _{in} } \right\rangle  = \sqrt {\frac{1}{2}} \left| {IW} \right\rangle  + \sqrt {\frac{1}{2}} \left| {NW} \right\rangle$. }}

In this example, the payoff of employer $A$ becomes negative, and that of worker $B$ is a relatively high value.

NEs: $\{ (p^* ,0)|p^*  \in [0,1]\}$.

$\bar \$ _A (p^* ,0) = -14$, $\bar \$ _B (p^* ,0) = 10$, $\forall p^* \in [0,1]$.

{\textbf{Example 6: $\left| {\psi _{in} } \right\rangle  = \sqrt {\frac{1}{2}} \left| {IS} \right\rangle  + \sqrt {\frac{1}{2}} \left| {NS} \right\rangle$. }}

Here, similar to Example 5, the payoff of employer $A$ is always negative, no matter what strategy he takes.

NEs: $\{ (p^* ,1)|p^*  \in [0,1]\}$.

$\bar \$ _A (p^* ,1) = -14$, $\bar \$ _B (p^* ,1) = 10$, $\forall p^* \in [0,1]$.

{\textbf{Example 7: $\left| {\psi _{in} } \right\rangle  = \sqrt {\frac{1}{4}} \left| {IW} \right\rangle  + \sqrt {\frac{1}{4}} \left| {IS} \right\rangle  + \sqrt {\frac{1}{4}} \left| {NW} \right\rangle  + \sqrt {\frac{1}{4}} \left| {NS} \right\rangle$. }}

In this scenario, the results are shown as follows.

NEs: $\{ (p^* ,q^* )|p^* ,q^*  \in [0,1]\}$.

$\bar \$ _A (p^* ,q^*) = 11$, $\bar \$ _B (p^* ,q^*) = 7.5$, $\forall p^*, q^* \in [0,1]$.

Obviously, this example presents very interesting results. In this setting of initial quantum state, the strategies of players do not affect the equilibrium of this quantum game. The joint payoff of $A$ and $B$, which equals to 18.5, is still lower than that of the NE of classical game.

\section{Summary}
To sum, we have extended the classical inspection game into a quantum version based on  Marinatto-Weber quantum game model. Different from the unique mixed-strategy NE of classical inspection game, quantizing this game and setting a suitable initial quantum state for the two-qubit system can lead to either pure-strategy NE or mixed-strategy NE. Moreover, compared with the classical inspection game, the quantum form has two main characteristics. At first, in the quantum inspection game, either employer or worker has ways to obtain more benefit than that derived from NE of classical inspection game. Another character is that the quantization can not bring Pareto improvement to the classical inspection game. In the quantum version, the employer or worker who wants to improve his benefit has to harm the interest of the other. The results show that there may exist irreconcilable conflict between employer and worker in the inspection game, which is different from prisoner's dilemma where Pareto efficiency can be carried out by using quantum strategies \cite{eisert1999quantum}. This work sheds a new insight to the inspection game and quantum game theory.

\begin{acknowledgments}
The work is partially supported by China Scholarship Council, National Natural Science Foundation of China (Grant No. 61174022), Specialized Research Fund for the Doctoral Program of Higher Education (Grant No. 20131102130002), R\&D Program of China (2012BAH07B01), National High Technology Research and Development Program of China (863 Program) (Grant No. 2013AA013801), the open funding project of State Key Laboratory of Virtual Reality Technology and Systems, Beihang University (Grant No.BUAA-VR-14KF-02).
\end{acknowledgments}

\bibliography{References}

\end{document}